\documentclass[]{spie}

\usepackage{amsmath,amsfonts,amssymb}
\usepackage{graphicx}
\usepackage[colorlinks=true, allcolors=blue]{hyperref}
\usepackage{tikz}
\usetikzlibrary{calc,arrows.meta}
\usepackage{subcaption}

\title{Simultaneous Dual-Plane Multi-Write-Spot Two-Photon Polymerization Using a Single Diffractive Optical Element}

\author[a]{Thomas Le Deun}
\author[a]{Joël Rovera}
\author[a]{Kevin Heggarty}

\affil[a]{IMT Atlantique, Brest, France}

\authorinfo{Further author information: (Send correspondence to T.L.D.)\\
T.L.D.: E-mail: thomas.ledeun@imt-atlantique.fr}

\begin{document}
\maketitle
\setlength{\emergencystretch}{3em}

Preprint submitted to SPIE Photonics Europe 2026. The final version will be published in the Proceedings of SPIE.

\begin{abstract}

The serial nature of two-photon polymerization (2PP) limits fabrication throughput. While diffractive optical elements (DOEs) can be used to generate multiple write spots in a single plane, three-dimensional structures still require sequential layer-by-layer fabrication. We demonstrate a dual-plane multi-spot 2PP approach using a single static DOE capable of generating two independent focal-spot arrays in distinct planes. This configuration enables simultaneous fabrication of two layers during continuous scanning while maintaining a simple scanning strategy compatible with woodpile structures. Using 29 write spots distributed across two planes separated by $1.8~\mu$m, we fabricate four-layer woodpile structures with an effective writing speed of $1~\mathrm{mm}^2$ in $90$ s. The results demonstrate that combining multi-write-spot parallelization with simultaneous multi-plane writing provides a powerful route to significantly increase 2PP fabrication throughput.

\end{abstract}


\section{INTRODUCTION}

Two-photon polymerization (2PP) is a well-established direct laser writing technique for fabricating complex three-dimensional microstructures with sub-micrometer resolution. However, throughput remains limited by the serial nature of the write process. In practice, fabrication throughput can be increased either by scanning a single focus faster or by parallelizing the exposure using multiple write spots. Increasing the scan speed is often the most power-efficient first option, since the required laser power typically increases sublinearly with scan speed, whereas in multi-focus approaches the total required power increases approximately in proportion to the number of simultaneously used spots \cite{kiefer2020sensitive}. Diffractive optical elements (DOEs) nevertheless provide an attractive route to further increase throughput once the achievable single-focus scan speed reaches practical limits and additional laser power is available. While DOE implementations usually focus on multi-spot parallelization within a single plane, multi-plane writing offers an additional degree of parallelization by reducing the number of field scans required at different Z positions while fabricating 3D structures without increasing mechanical complexity.

Here, we demonstrate dual-plane 2PP using a single DOE capable of generating two independent focal-spot distributions located in distinct planes. By combining in-plane parallelization with simultaneous two-plane writing, the proposed approach enables a significant increase in fabrication throughput while preserving a simple scanning strategy, leading to a potential throughput increase exceeding one order of magnitude.

\section{DOE-based photoplotter system}

\begin{figure}[ht]
\centering
\includegraphics[width=0.6\textwidth]{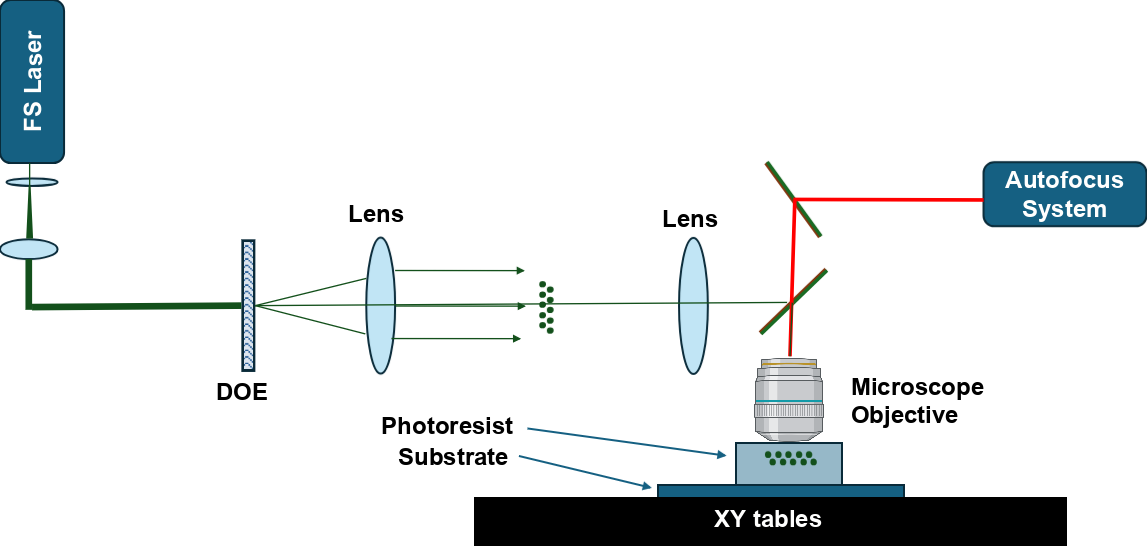}
\caption{Schematic of the experimental setup used for dual-plane two-photon polymerization.}
\label{fig:experimental_setup}
\end{figure}

The experiments were carried out with a femtosecond two-photon polymerization setup schematically shown in Fig.~\ref{fig:experimental_setup}. The beam delivered by a frequency-doubled femtosecond laser source (NKT Photonics aeroPULSE FS10, 515 nm, operated at 1 MHz, 5W power available) was expanded and sent onto a static diffractive optical element fabricated in-house by photolithography using our classical one-photon photoplotter \cite{kessels2007versatile}. The DOE was relayed to the rear aperture of the microscope objective (Zeiss oil-immersion objective, 40x, 1.4 NA) using a 4f imaging system. In this configuration, the focal intensity distribution formed in the photoresist corresponds to the Fourier transform of the DOE phase pattern. Imaging the DOE onto the objective pupil reduces vignetting and helps preserve the quality of the generated multi-spot distribution over the usable writing field. In addition, the intermediate Fourier plane of the 4f system provides physical access to the diffracted spot array, which can be useful for spatial filtering of unwanted diffraction orders and can thereby greatly simplify DOE design and fabrication.

The write spots were focused into the photoresist through the microscope objective while the sample was translated using air-bearing motorized XY stages (Physik Instrumente A-311.BB1) combined with a Z axis (Physik Instrumente V-574). A differential confocal autofocus system \cite{yang2023autofocus} was used to detect the glass–resist interface and to establish a topography map of this interface prior to fabrication. In the present work, the fabricated DOE generated a dual-plane multi-spot distribution composed of two diagonal sub-arrays located in different planes. During continuous scanning, each focal spot produced a polymerized line, so that two layers could be written simultaneously in a single pass. By successive scans along the X and Y directions, this configuration enabled the direct fabrication of woodpile-type structures.

The design strategy used to generate the two focal-spot distributions in distinct planes is described in the following section.

\section{DOE DESIGN AND FOCAL-PLANE SEPARATION STRATEGY}

The diffractive optical element was designed using in-house IFTA-based algorithms \cite{ripoll2004review} to generate two independent spot distributions located in distinct planes. A diagonal geometry was deliberately chosen for the spot arrays so that the same writing strategy can be applied when scanning either along the X or the Y direction. This symmetry enables the fabrication of orthogonal layers using identical scanning conditions, which is particularly convenient for the fabrication of woodpile-type structures.

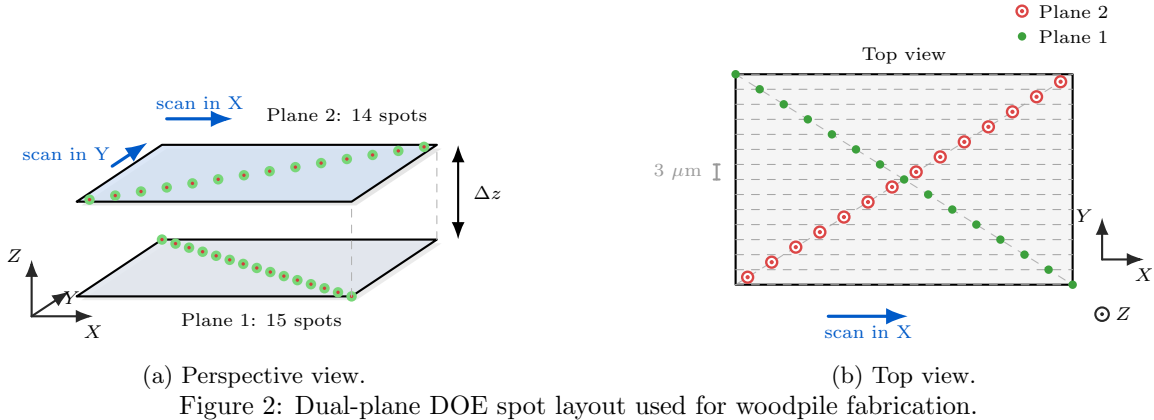
\begin{figure}[ht]
\centering

\begin{subfigure}[t]{0.48\textwidth}
\centering
\begin{tikzpicture}[
    scale=0.63,
    line join=round,
    line cap=round,
    >=Latex,
    every node/.style={font=\scriptsize}
]

\definecolor{planetop}{RGB}{210,225,245}
\definecolor{planebot}{RGB}{225,230,238}
\definecolor{spotouter}{RGB}{120,215,120}
\definecolor{spotinner}{RGB}{210,60,60}
\definecolor{axiscol}{RGB}{40,40,40}
\definecolor{scancol}{RGB}{0,90,200}

\coordinate (A) at (0.8,0.7);
\coordinate (B) at (6.6,0.7);
\coordinate (C) at (8.4,1.9);
\coordinate (D) at (2.6,1.9);

\coordinate (dz) at (0,2.0);
\coordinate (E) at ($(A)+(dz)$);
\coordinate (F) at ($(B)+(dz)$);
\coordinate (G) at ($(C)+(dz)$);
\coordinate (H) at ($(D)+(dz)$);

\fill[black!8]  ($(A)+(0.10,-0.08)$) -- ($(B)+(0.10,-0.08)$) -- ($(C)+(0.10,-0.08)$) -- ($(D)+(0.10,-0.08)$) -- cycle;
\fill[black!10] ($(E)+(0.10,-0.08)$) -- ($(F)+(0.10,-0.08)$) -- ($(G)+(0.10,-0.08)$) -- ($(H)+(0.10,-0.08)$) -- cycle;

\fill[planebot, opacity=0.85] (A) -- (B) -- (C) -- (D) -- cycle;
\draw[thick] (A) -- (B) -- (C) -- (D) -- cycle;

\fill[planetop, opacity=0.78] (E) -- (F) -- (G) -- (H) -- cycle;
\draw[thick] (E) -- (F) -- (G) -- (H) -- cycle;

\foreach \i in {0,...,14} {
    \pgfmathsetmacro{\tbot}{\i/14}
    \path ($(D)!\tbot!(B)$) coordinate (Pbot\i);
    \fill[spotouter] (Pbot\i) circle (0.11);
    \fill[spotinner] (Pbot\i) circle (0.036);
}

\foreach \i in {0,...,13} {
    \pgfmathsetmacro{\ttop}{(\i+0.5)/14}
    \path ($(E)!\ttop!(G)$) coordinate (Ptop\i);
    \fill[spotouter] (Ptop\i) circle (0.11);
    \fill[spotinner] (Ptop\i) circle (0.036);
}

\draw[<->, thick] (8.85,1.9) -- (8.85,3.9);
\node[right] at (8.98,2.9) {$\Delta z$};

\draw[dashed, gray!60] (B) -- (F);
\draw[dashed, gray!60] (C) -- (G);

\node[align=center] at (6.5,4.5) {Plane 2: 14 spots};
\node[align=center] at (4.7,0.18) {Plane 1: 15 spots};

\draw[->, very thick, scancol] (2.7,4.45) -- (4.1,4.45);
\node[above, scancol] at (3.4,4.45) {scan in X};

\draw[->, very thick, scancol] (1.55,3.45) -- (2.3,3.95);
\node[above left, scancol] at (1.7,3.38) {scan in Y};

\coordinate (O) at (-0.15,0.28);

\draw[->, thick, axiscol] (O) -- ++(1.3,0.0);
\node[below] at (1.15,0.28) {$X$};

\draw[->, thick, axiscol] (O) -- ++(0.8,0.53);
\node[below right] at (0.32,1.0) {$Y$};

\draw[->, thick, axiscol] (O) -- ++(0,1.2);
\node[left] at (-0.15,1.55) {$Z$};

\end{tikzpicture}
\caption{Perspective view.}
\label{fig:dualplane_schematic_a}
\end{subfigure}
\hspace{0.01\textwidth}
\begin{subfigure}[t]{0.48\textwidth}
\centering
\begin{tikzpicture}[
    scale=0.56,
    line join=round,
    line cap=round,
    >=Latex,
    every node/.style={font=\scriptsize}
]

\definecolor{topspots}{RGB}{220,70,70}
\definecolor{botspots}{RGB}{60,160,60}
\definecolor{scancol}{RGB}{0,90,200}
\definecolor{axiscol}{RGB}{40,40,40}

\coordinate (A) at (0,0);
\coordinate (B) at (8,0);
\coordinate (C) at (8,5);
\coordinate (D) at (0,5);

\fill[gray!8] (A) rectangle (C);
\draw[thick] (A) rectangle (C);

\draw[dashed, gray!60] (D) -- (B);
\draw[dashed, gray!60] (A) -- (C);

\foreach \i in {0,...,14} {
    \pgfmathsetmacro{\tbot}{\i/14}
    \pgfmathsetmacro{\ybot}{5 - 5*\tbot}
    \path ($(D)!\tbot!(B)$) coordinate (Pbot\i);
    \draw[dashed, gray!70] (0,\ybot) -- (8,\ybot);
    \fill[botspots] (Pbot\i) circle (0.10);
}

\foreach \i in {0,...,13} {
    \pgfmathsetmacro{\ttop}{(\i+0.5)/14}
    \path ($(A)!\ttop!(C)$) coordinate (Ptop\i);
    \draw[line width=0.9pt, topspots, fill=white] (Ptop\i) circle (0.13);
    \fill[topspots] (Ptop\i) circle (0.05);
}

\node[align=center] at (4,5.45) {Top view};

\draw[->, very thick, scancol] (2.2,-0.75) -- (4.1,-0.75);
\node[below, scancol] at (3.15,-0.75) {scan in X};

\pgfmathsetmacro{\yA}{5 - 5*(6/14)}
\pgfmathsetmacro{\yB}{5 - 5*(7/14)}

\draw[gray!80, thick] (-0.52,\yA) -- (-0.38,\yA);
\draw[gray!80, thick] (-0.52,\yB) -- (-0.38,\yB);

\draw[gray!80, thick] (-0.45,\yA) -- (-0.45,\yB);

\node[left, gray!80] at (-0.58,{(\yA+\yB)/2}) {$3~\mu$m};

\coordinate (O) at (8.7,0.6);

\draw[->, thick, axiscol] (O) -- ++(1.0,0);
\node[below] at (9.7,0.6) {$X$};

\draw[->, thick, axiscol] (O) -- ++(0,1.0);
\node[left] at (8.7,1.65) {$Y$};

\draw[thick, axiscol] (8.7,-0.7) circle (0.16);
\fill[axiscol] (8.7,-0.7) circle (0.045);
\node[right] at (8.8,-0.7) {$Z$};

\fill[botspots] (6.8,5.9) circle (0.10);
\node[right] at (6.98,5.9) {Plane 1};

\draw[line width=0.9pt, topspots, fill=white] (6.8,6.5) circle (0.13);
\fill[topspots] (6.8,6.5) circle (0.05);
\node[right] at (6.98,6.5) {Plane 2};

\end{tikzpicture}
\caption{Top view.}
\label{fig:dualplane_schematic_b}
\end{subfigure}

\caption{Dual-plane DOE spot layout used for woodpile fabrication. }
\label{fig:dualplane_combined}
\end{figure}

As illustrated in Fig.~\ref{fig:dualplane_combined}, in the first plane, the DOE generates a diagonal array of 15 focal spots oriented at $45^\circ$ relative to the scan direction. During continuous scanning along the X direction, these spots produce parallel polymerized lines with a spacing of 3 µm. In the second plane, the DOE generates 14 spots oriented along the opposite diagonal direction and interleaved between the 15 focal spots. To introduce a focal-plane separation between the two spot distributions, a quadratic Fresnel phase term was added
to the second phase mask. 

In our optical configuration, the 4f relay images the DOE phase mask onto the rear pupil of the microscope objective, so that the quadratic Fresnel phase term can be treated as an additional thin lens located at the objective pupil. Using a thin-lens approximation, the vergence of this virtual lens is then combined with that of the microscope objective. Taking into account propagation in the immersion medium of refractive index $n$, this leads to an estimated focal-plane separation of
\begin{equation}
\Delta z = n\,\frac{f_{\mathrm{obj}}^2}{f+f_{\mathrm{obj}}}
\approx n\,\frac{f_{\mathrm{obj}}^2}{f},
\end{equation}
where $f$ is the focal length of the Fresnel phase term, $f_{\mathrm{obj}}$ is the effective focal length of the objective, and the approximation holds for $f\gg f_{\mathrm{obj}}$. In our case ($f=12$ m, $f_{\mathrm{obj}}\approx 4.125$ mm, $n\approx1.5$), this gives a focal-plane separation of $2.1~\mu$m.

\begin{figure}[ht]
\centering
\includegraphics[width=0.3\textwidth]{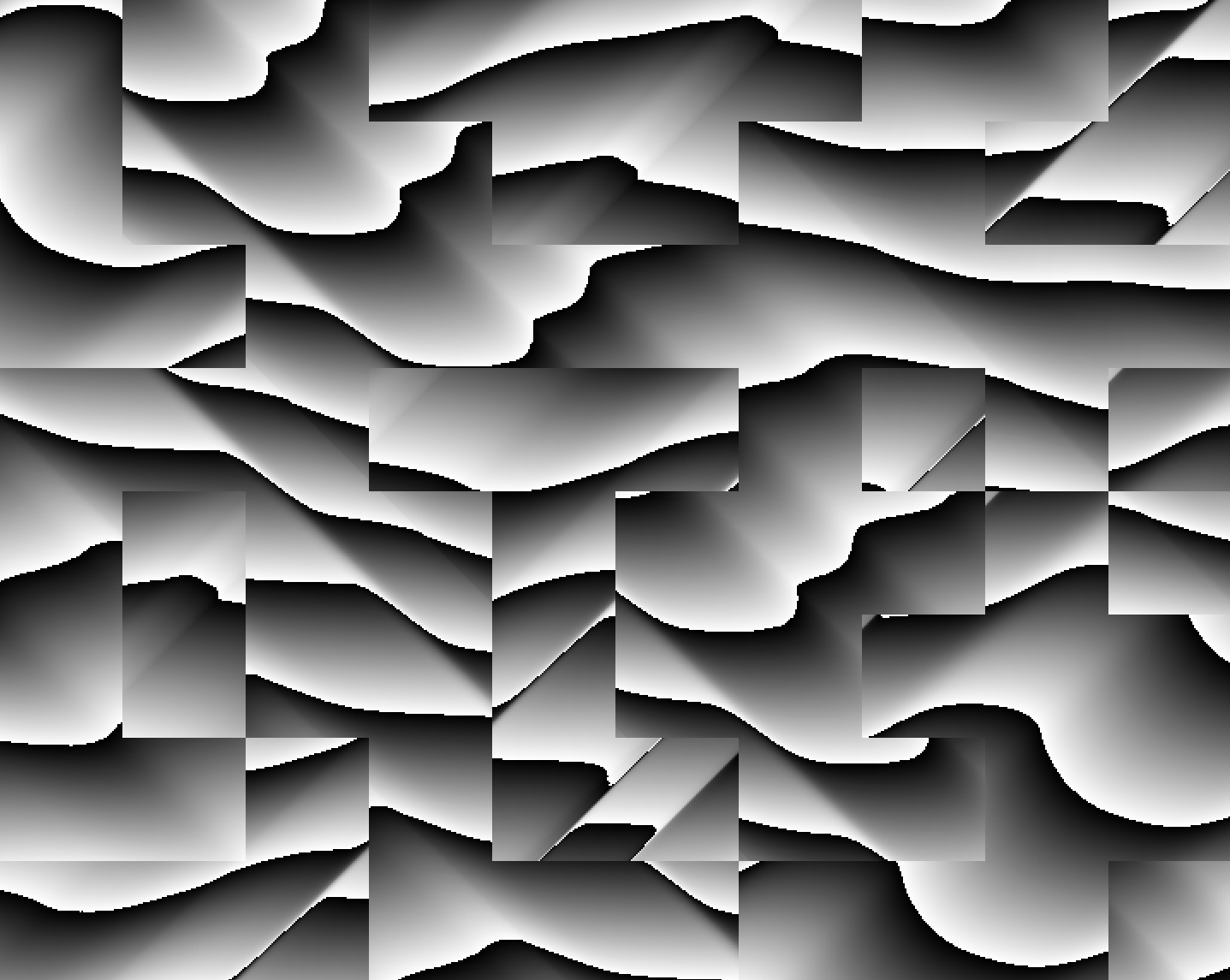}
\caption{700x700 pixels portion of the composite DOE phase mask obtained by random interleaving of 72x72 pixel blocks from the two independently computed phase masks.}
\label{fig:doe_phase_mask}
\end{figure}

The final composite DOE was generated by randomly interleaving $72\times72$ pixel blocks extracted from the two independently computed phase masks. This stochastic spatial multiplexing avoids the periodic artifacts that would arise from regular checkerboard segmentation and suppresses unwanted diffraction orders while preserving the functionality of the phase mask. An example of the resulting composite phase mask is shown in Fig.~\ref{fig:doe_phase_mask}, illustrating the random spatial interleaving of the two phase distributions.

\section{RESULTS}

To demonstrate the capabilities of the dual-plane writing configuration, we fabricated woodpile structures. The fabrication sequence consisted of 10 scan passes along the X direction followed by 10 scan passes along the Y direction, with a respective distance between scans of $45~\mu$m, corresponding to the total width of the write spots in the focal plane.
The focal-plane separation was experimentally estimated by writing parallel lines along the X direction while applying different Z offsets relative to the glass–resist interface previously identified by the autofocus system. The offsets for which the 15-spot and 14-spot arrays produced sharply focused lines at the interface were then compared. This procedure gave an experimental focal-plane separation of $1.8~\mu$m, in relatively good agreement with the predicted value of $2.1~\mu$m given the simplicity of the thin-lens approximation.
Between these two scan sequences, the focus position was shifted by an offset equal to $\Delta z/2$, corresponding to $0.9~\mu$m in our experiment. Because the DOE writes two layers simultaneously, this procedure produces four distinct layers of polymerized lines forming a woodpile structure. The fabrication was performed with a scan speed of $5~\mathrm{mm/s}$ and a laser output power of $300~\mathrm{mW}$. With 29 simultaneous write spots, this corresponds to an effective writing speed of $145~\mathrm{mm/s}$ across two planes written simultaneously. Considering the approximately $50~\mu$m $\times$ $50~\mu$m writing footprint of the DOE, a $1~\mathrm{mm}^2$ four-layer woodpile structure can be fabricated in about $90$ seconds.

\begin{figure}[ht]
\centering

\begin{subfigure}[t]{0.41\textwidth}
\centering
\includegraphics[width=\textwidth]{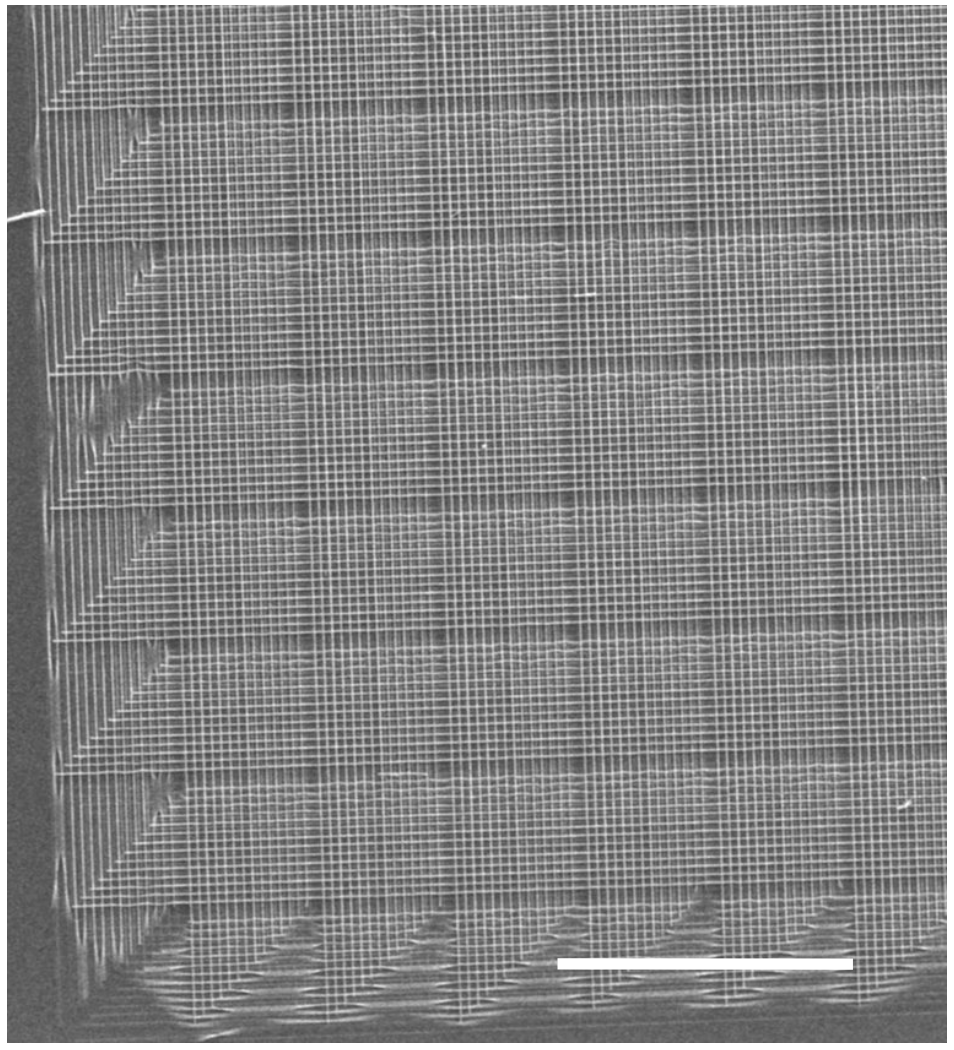}
\caption{Large-area SEM view ($100~\mu$m scale bar).}
\label{fig:woodpile_large}
\end{subfigure}
\hspace{0.02\textwidth}
\begin{subfigure}[t]{0.438\textwidth}
\centering
\includegraphics[width=\textwidth]{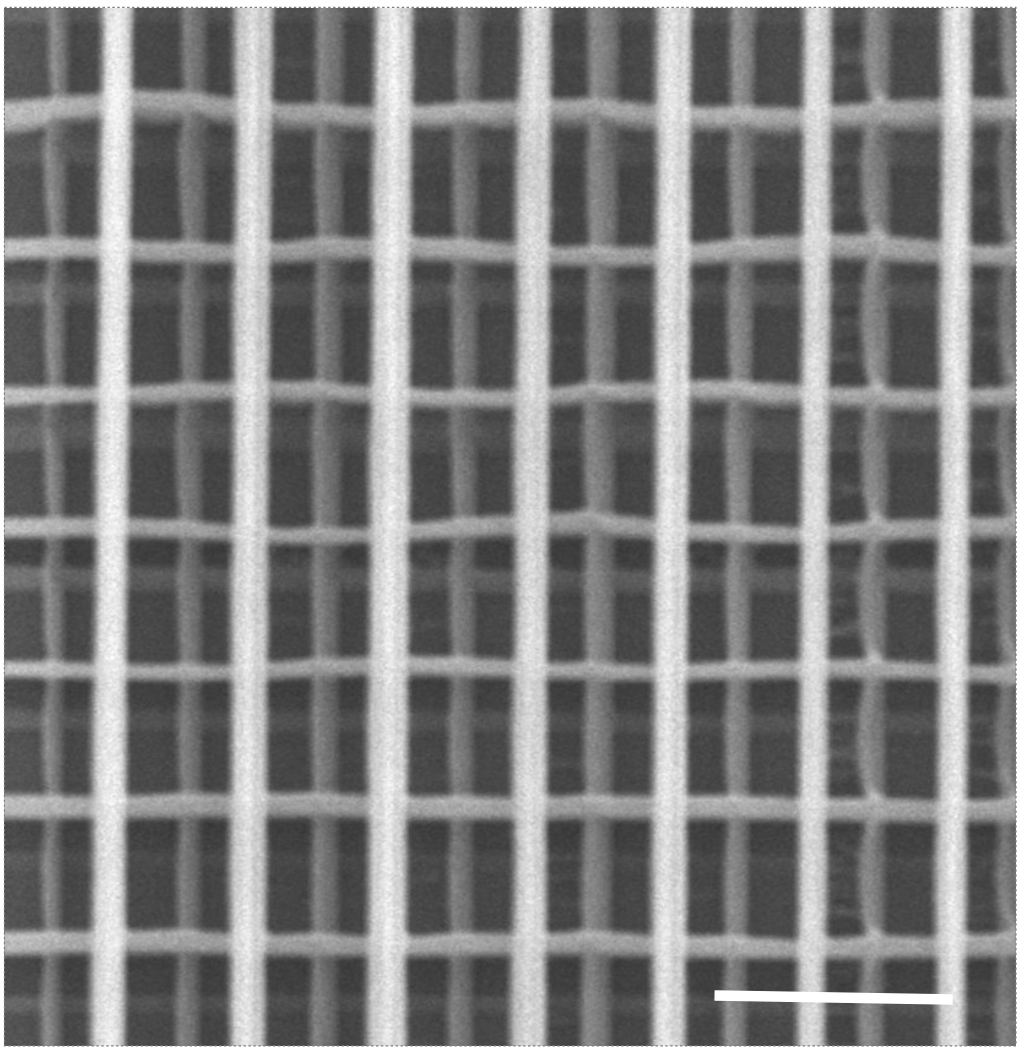}
\caption{Higher-magnification SEM image showing the layered woodpile structure ($5~\mu$m scale bar).}
\label{fig:woodpile_close}
\end{subfigure}

\caption{Scanning electron microscope (SEM) images of the fabricated woodpile structures produced using the dual-plane multi-spot 2PP approach.}
\label{fig:woodpile_sem}
\end{figure}

Scanning electron microscope (SEM) images confirm the presence of four well-separated layers forming the expected woodpile structure as illustrated in Fig.~\ref{fig:woodpile_sem}.

These results illustrate the potential of combining multi-write-spot parallelization with multi-plane writing to increase fabrication throughput in two-photon polymerization. Further throughput improvements could be achieved by increasing the number of write spots generated by the DOE. In practice, we observed that up to approximately 200 spots could be generated before reaching the damage threshold of the microscope objective, as the required laser power scales roughly linearly with the number of foci. The approach could also be extended to multiplex more than two planes. In this case, the main limitation would arise from the contrast between the focal spot distributions in the different planes, which will decrease as additional phase components are combined within a single DOE.

\acknowledgments
This work has received funding from the European Union’s Horizon Europe research and innovation programme under grant agreement nº 101091644. UK participants in Horizon Europe Project FABulous are supported by UKRI grant nº 10062385 (MODUS). The authors acknowledge financial support from French National Research Agency (ANR) through the NANOSHAPE project (ANR-23-INDF-0003). The authors thank Anaïs Calvez for fabrication and characterization, and Valeriia Sedova for insightful discussions on proximity effect compensation.

\bibliographystyle{spiebib}
\bibliography{report}

\end{document}